\documentclass[twocolumn,showpacs,amsfonts,aps,prc,floatfix,superscriptaddress]{revtex4-1}

\usepackage[colorlinks=true, pdfstartview=FitV, linkcolor=red, citecolor=blue, urlcolor=blue]{hyperref}
\usepackage{amsmath}
\usepackage{bm}
\usepackage{epsfig}
\usepackage{color}
\usepackage{epstopdf}
\usepackage{graphicx}
\usepackage[normalem]{ulem}

\begin{document}

\title{
Four-dimensional QCD equation of state with multiple chemical potentials
}

\author{Akihiko Monnai}
\affiliation{Department of General Education, Faculty of Engineering, \\
Osaka Institute of Technology, Osaka 535-8585, Japan}

\author{Gr\'{e}goire Pihan}
\affiliation{Department of Physics and Astronomy, Wayne State University, Detroit, Michigan, USA}

\author{Bj\"orn Schenke}
\affiliation{Physics Department, Brookhaven National Laboratory, Upton, NY 11973, USA}

\author{Chun Shen}
\affiliation{Department of Physics and Astronomy, Wayne State University, Detroit, Michigan, USA}
\affiliation{RIKEN BNL Research Center, Brookhaven National Laboratory, Upton, NY 11973, USA}

\date{\today}

\begin{abstract}
We construct a four-dimensional version of the equation of state (EoS) model \textsc{neos}, \textsc{neos-4d}, as a function of the temperature and chemical potentials of baryon, electric charge, and strangeness for the hot and dense QCD matter created in relativistic nuclear collisions. This EoS enables multiple conserved charge current evolution in a relativistic fluid. Input from Lattice QCD simulations and a hadron resonance gas model is considered for constructing the equation of state. We investigate its applicability to the relativistic hydrodynamic description of nuclear collisions and present a method for efficient numerical implementation. 
\end{abstract}

\pacs{25.75.-q, 21.65.Qr, 12.38.Mh}

\maketitle

\section{Introduction}
\label{sec1}

The phase structure of quantum chromodynamics (QCD) is a topic of fundamental importance in particle and nuclear physics. The state of a system near thermal equilibrium is characterized by the equation of state. With the advent of realistic Lattice QCD simulations, the QCD equation of state at finite temperatures has become available with high precision. It has been found that the (2+1)-flavor QCD system has a crossover-type transition \cite{Brown:1990ev,AliKhan:2000wou,Aoki:2006we} from the hadronic phase to the quark-gluon plasma (QGP) phase at temperatures around 155-160~MeV in the limit of vanishing chemical potentials \cite{Borsanyi:2013bia, Bazavov:2014pvz,Bollweg:2022fqq}.

Thermodynamic properties of the QCD system at finite chemical potentials, on the other hand, cannot be computed directly because of the fermion sign problem of Lattice QCD simulations \cite{deForcrand:2010ys}. Model-based approaches suggest the existence of a rich structure of the phase diagram at finite densities \cite{Fukushima:2010bq,Du:2024wjm}, including a critical point which marks where the cross over turns into a first-order phase transition between the QGP and hadronic phases \cite{Asakawa:1989bq}. Elucidation of the QCD equation of state over a wide range of temperatures and chemical potentials is an important topic in various fields of theoretical physics, including high-energy nuclear collisions and compact stars \cite{Baym:2017whm,Fujimoto:2019hxv,Dexheimer:2020zzs,Komoltsev:2021jzg,MUSES:2023hyz}.

Relativistic nuclear collisions at the Relativistic Heavy Ion Collider (RHIC) at Brookhaven National Laboratory \cite{Adcox:2004mh,Adams:2005dq,Back:2004je,Arsene:2004fa} and the CERN Large Hadron Collider (LHC) \cite{Aamodt:2010pa,ATLAS:2011ah,Chatrchyan:2012wg} provide a prodigious amount of opportunities to study deconfined quark matter. One of the most important discoveries is the revelation that the QCD matter produced in high energy collisions behaves as a fluid with extremely small viscosity. The relativistic hydrodynamic model has become an instrumental tool for simulating the dynamical evolution of the bulk QCD matter and for predicting the produced particle distributions observed in heavy ion collider experiments \cite{Kolb:2000fha,Schenke:2010rr}. Lattice QCD-based equations of state have been used in recent hydrodynamic models, and have proven successful in reproducing the experimental data at high energies where the baryon chemical potential of the system is negligible. 

The beam energy scan (BES) program at the RHIC \cite{STAR:2010vob} as well as the NA61/SHINE experiment at the CERN Super Proton Synchrotron (SPS) have been performed to obtain experiment driven insight into the phase structure and finite-density properties of QCD. Similar experiments are being planned at various facilities including the GSI Facility for Antiproton and Ion Research (FAIR) and JAEA/KEK Japan Proton Accelerator Research Complex (J-PARC). The hydrodynamic model, supplemented with a finite-density equation of state \cite{Huovinen:2009yb,Moreland:2015dvc,Parotto:2018pwx,Monnai:2019hkn,Noronha-Hostler:2019ayj,Auvinen:2020mpc,Monnai:2021kgu,Kahangirwe:2024cny,Plumberg:2024leb}, should play a pivotal role in describing the dense quark matter created in nuclear collisions and in extracting microscopic information on QCD from the experimental data. 

The conserved charges in the strongly-interacting matter created in nuclear collisions are net baryon (B), electric charge (Q), and strangeness (S) \cite{Werner:2010aa,Monnai:2019hkn,Noronha-Hostler:2019ayj,Aryal:2020ocm,Karthein:2021nxe,Schafer:2021csj,Monnai:2021kgu}. The strangeness neutrality condition $n_S = 0$ and the fixed charge-to-baryon ratio $n_Q/n_B \sim 0.4$ are often employed to simulate collisions of heavy nuclei such as $^{197}_{\ 79}$Au and $^{208}_{\ 82}$Pb. The equation of state then effectively reduces to two dimensions. The assumption of uniform charge ratios, however, does not hold when protons and neutrons are distinguished or when local fluctuations and/or diffusion processes are considered. Also, collisions involving light nuclei would have different values of the charge-to-baryon ratio. Thus, it is critical to develop a four-dimensional equation of state in $T, \mu_B, \mu_Q,$ and $\mu_S$ for comprehensive studies of the experimentally-created QCD matter at finite densities using hydrodynamic models \cite{Pihan:2023dsb,Pihan:2024lxw,Plumberg:2023vkw,Plumberg:2024leb}. 

We consider the equation of state model \textsc{neos} \cite{Monnai:2019hkn} and extend it to treat multiple chemical potentials without constraints on the charge ratios. We note that a similar model using different methods can be found in Ref.~\cite{Plumberg:2024leb}. In our study, the pressure of state-of-the-art Lattice QCD simulations based on the Taylor expansion method \cite{Gavai:2001fr,Allton:2002zi} is matched to that of the hadron resonance gas model in the vicinity of the crossover \cite{Borsanyi:2020fev}. The inclusion of the critical point \cite{Nonaka:2004pg,Karthein:2021nxe,Grefa:2021qvt} is left for future work. We further develop an efficient method to utilize this QCD equation of state in numerical simulations of the hydrodynamic model of relativistic nuclear collisions. 

This paper is organized as follows. In Sec.~\ref{sec2}, we present the four-dimensional version of our QCD equation of state model and discuss the region of phase space explored by nuclear collisions. Numerical implementation into the hydrodynamic modeling, conservation at particlization, and comparison of equations of state with different hadronic particle contents are discussed in Sec.~\ref{sec3}. Sec.~\ref{sec4} is devoted to conclusions. We use the natural units $c = \hbar = k_B = 1$ and the mostly-minus metric $g^{\mu \nu} = \mathrm{diag}(+,-,-,-)$.

\section{The equation of state}
\label{sec2}

We develop \textsc{neos-4d}, a four-dimensional equation of state with the four dimensions spanned by temperature and multiple chemical potentials. The equation of state and susceptibilities of Lattice QCD simulations at vanishing densities \cite{Borsanyi:2011sw,Bellwied:2015lba,Borsanyi:2018grb,Bazavov:2012jq, Ding:2015fca, Bazavov:2017dus} are used to construct a finite-density equation of state in the Taylor expansion method \cite{Gavai:2001fr,Allton:2002zi} at high temperatures. Since the expansion in terms of fugacities is not reliable at lower temperatures \cite{Karsch:2010hm}, the hadron resonance gas model, an effective model that accounts for stable hadrons and meta-stable resonances, is employed for describing the hadronic phase. This prescription also allows smooth matching of the hydrodynamic and hadronic transport models via particlization.

The light quarks $u$, $d$, and $s$ are considered to constitute the quark-gluon plasma because heavy quarks would not be in equilibrium with the medium at temperatures reached in current relativistic nuclear colliders. Consequently, the hadrons that have $u$, $d$, and/or $s$ as valence quarks are taken into consideration.

\subsection{Construction}

We begin by reviewing the construction of the \textsc{neos} model \cite{Monnai:2019hkn,Monnai:2021kgu}. 
The crossover-type equation of state is estimated by the smooth connection of the pressures of the lattice QCD simulations and of the hadron resonance gas model. They are connected in the vicinity of the crossover as 
\begin{align}
P &= \frac{1}{2}\bigg(1- \tanh \frac{T-T_c}{\Delta T_c}\bigg) P_{\mathrm{had}} \nonumber \\
&+ \frac{1}{2}\bigg(1+ \tanh \frac{T-T_c}{\Delta T_c}\bigg) P_{\mathrm{lat}} , \label{eq:econ}
\end{align}
where $T_c$ and $\Delta T_c$ are the connecting temperature and width, respectively. $P$ approaches $P_{\mathrm{had}}$ towards lower temperatures and $P_\mathrm{lat}$ towards higher temperatures. 

The pressure in the hadron resonance gas model is 
\begin{equation}
P_\mathrm{had} = \pm T \sum_i \int \frac{g_i d^3p}{(2\pi)^3} \ln [1 \pm e^{-(E_i-\mu_i)/T} ] ,\label{eq:P_had}
\end{equation}
where $i$ is the index for particle species, $g_i$ is the degeneracy, $E_i = \sqrt{p^2 + m_i^2}$ is the energy, and $m_i$ is the particle mass. $\mu_i = B_i \mu_B + Q_i \mu_Q + S_i \mu_S$ is the hadronic chemical potential where $B_i$, $Q_i$, and $S_i$ are the quantum numbers associated with baryon, electric charge, and strangeness, respectively. The signs are $+$ for fermions and $-$ for bosons.

The lattice QCD pressure in the Taylor expansion method is
\begin{equation}
\frac{P_\mathrm{lat}}{T^4} = \frac{P_0}{T^4} + \sum_{l,m,n} \frac{\chi^{B,Q,S}_{l,m,n}}{l!m!n!} \bigg( \frac{\mu_B}{T} \bigg)^{l}  \bigg( \frac{\mu_Q}{T} \bigg)^{m}  \bigg( \frac{\mu_S}{T} \bigg)^{n}, 
\label{Psus}
\end{equation}
where $P_0$ and $\chi^{B,Q,S}_{l,m,n}$ are the pressure and the $(l+m+n)$-th order susceptibility defined at zero chemical potentials. 
One has to be careful that the results that involve the regions with large $\mu_B/T (>3)$, $\mu_S/T$ or $\mu_Q/T$ should be considered extrapolations because the Taylor expansion Eq.~\eqref{Psus} is not well-defined there.
All the diagonal and off-diagonal susceptibilities up to the fourth order are used for the construction. In addition, some of the sixth order susceptibilities, $\chi_{6}^B$, $\chi_{51}^{BQ}$, and $\chi_{51}^{QS}$, which would contribute most to $n_B$, $n_Q$, and $n_S$, respectively, are introduced to preserve thermodynamic consistency, which requires that the state variables are smooth and monotonic.

\subsection{Numerical simulations}

\begin{figure}[tb]
\includegraphics[width=3.4in]{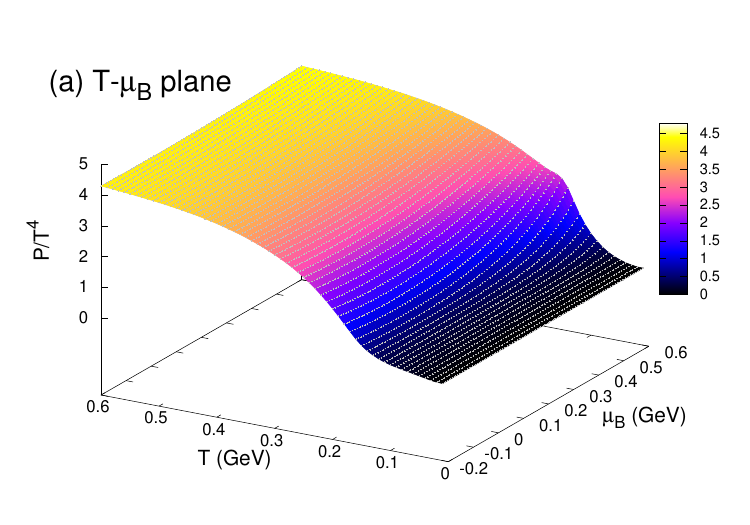}
\includegraphics[width=3.4in]{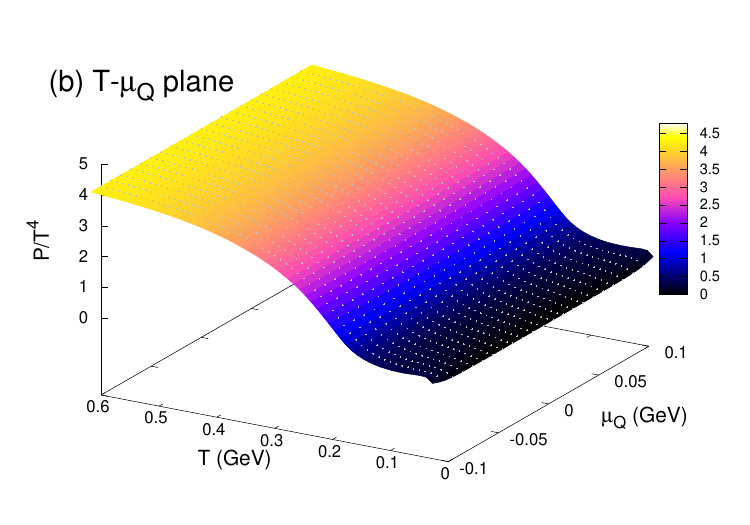}
\includegraphics[width=3.4in]{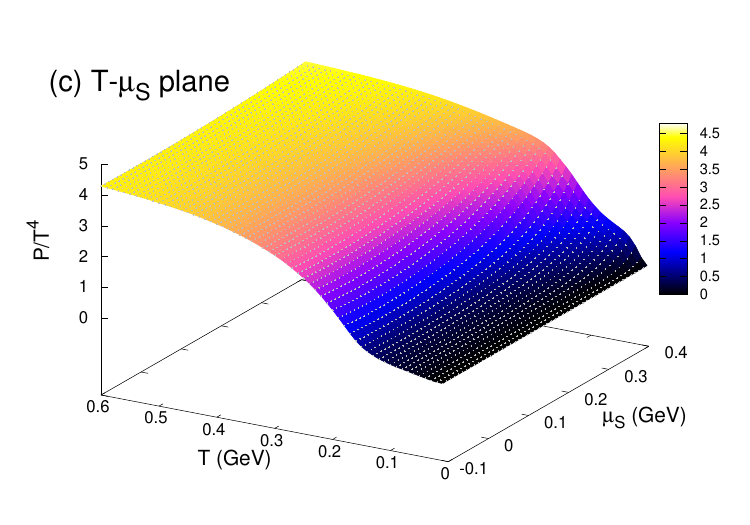}
\caption{(Color online) The dimensionless pressure $P/T^4$ as functions of (a) $T$ and $\mu_B$ at $\mu_S = \mu_Q = 0$, (b) $T$ and $\mu_Q$ at $\mu_B = \mu_S = 0$, and (c) $T$ and $\mu_S$ at $\mu_B = \mu_Q = 0$.}
\label{fig:pressure}
\end{figure}

\begin{figure}[tb]
\includegraphics[width=3.4in]{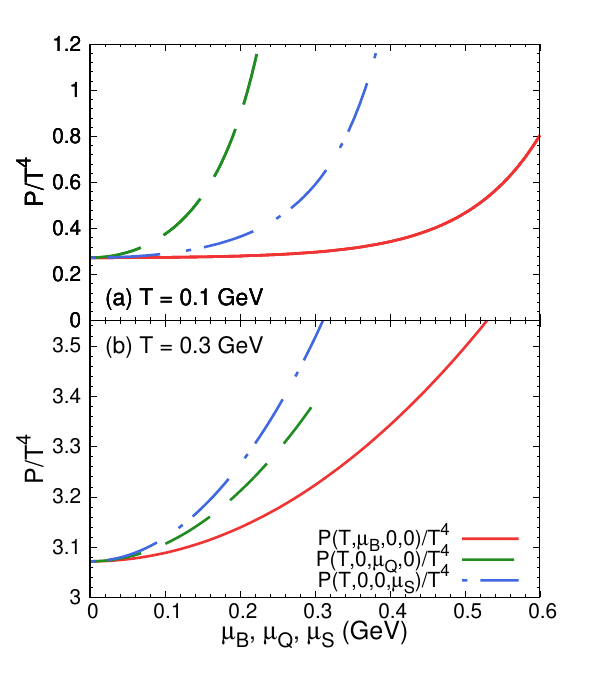}
\caption{(Color online) Dependence of $P/T^4$ on chemical potentials $\mu_B$, $\mu_Q$, and $\mu_S$ at (a) $T = 0.1$ GeV and (b) $T = 0.3$ GeV.}
\label{fig:mu_slices}
\end{figure}

We calculate the four-dimensional QCD equation of state numerically. The hadron resonances from the particle data group with mass up to 2 GeV are taken into account for the hadronic phase. The zero-density equation of state \cite{Bazavov:2014pvz} and susceptibilities up to the fourth order \cite{Bazavov:2012jq, Ding:2015fca, Bazavov:2017dus, Sharma} from Lattice QCD simulations are employed in the QGP phase. The aforementioned sixth-order susceptibilities are parameterized in accordance with Ref.~\cite{Monnai:2019hkn}. The connecting temperature is 
\begin{equation}
T_c(\mu_B) = a - d \, (b \mu_B^2 + c \mu_B^4), \label{eq:tc}
\end{equation}
where $a = 0.16\,\mathrm{GeV}$, $b = 0.139\,\mathrm{GeV}^{-1}$, $c = 0.053\,\mathrm{GeV}^{-3} $, and $d = 0.4$, whose form is motivated by a criterion for the chemical freeze-out \cite{Cleymans:2005xv}. The dependencies on the charge and strangeness chemical potentials are assumed to be small. The connecting width is chosen as $\Delta T_c = 0.1T_c (0)$.

The dimensionless pressure $P/T^4$ is obtained as a function of the temperature and three chemical potentials. Figure~\ref{fig:pressure} shows $P/T^4$ on three different two-dimensional slices of the four-dimensional space: (a) the $T$-$\mu_B$ plane at $\mu_Q=\mu_S=0$, (b) the $T$-$\mu_Q$ plane at $\mu_B=\mu_S=0$, and (c) the $T$-$\mu_S$ plane at $\mu_B=\mu_Q=0$. Negative baryon chemical potential regions are shown as the quantity can become negative owing to local fluctuation or diffusion in nuclear collisions. For clarity, slices of $P/T^4$ at the fixed temperatures of $T=0.1$ GeV and $0.3$ GeV as functions of $\mu_B$, $\mu_Q$, or $\mu_S$ are shown in Fig.~\ref{fig:mu_slices}. One can see that the pressure increases monotonically as the absolute value of the chemical potentials increases both in the hadronic and QGP phases. The dimensionless pressure is more sensitive to the change in $\mu_Q$ than that in $\mu_S$ or $\mu_B$ in the hadronic phase because the lightest hadrons that carry electric charge, strangeness, and net baryon charge are pions, kaons, and nucleons, respectively. The lighter these particles masses, the more they contribute to the pressure, explaining the faster rise as a function of $\mu_Q$ and $\mu_S$ compared with $\mu_B$. 
In the QGP phase, on the other hand, the pressure is most sensitive to $\mu_S$ which can be understood from the ideal parton gas picture \cite{Monnai:2019hkn}.

\begin{figure}[tb]
\includegraphics[width=3.4in]{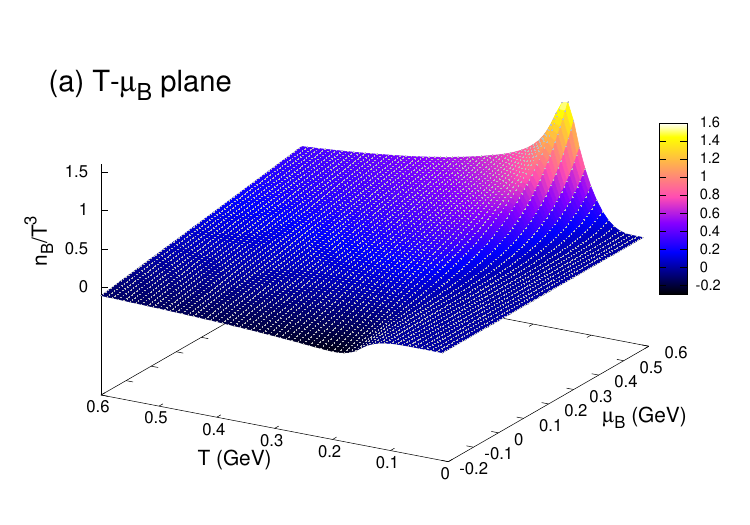}
\includegraphics[width=3.4in]{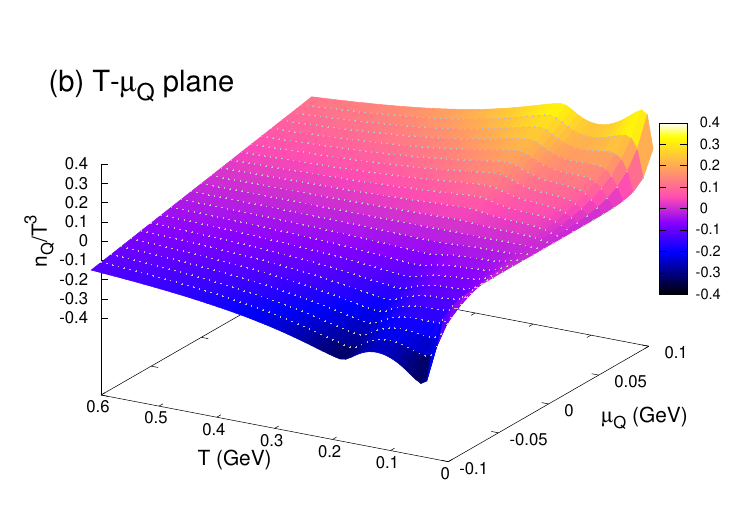}
\includegraphics[width=3.4in]{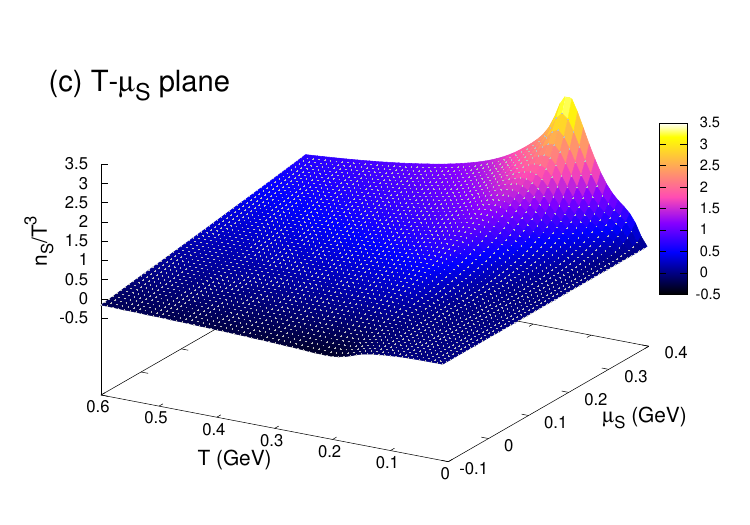}
\caption{(Color online) (a) $n_B/T^3$ as function of $T$ and $\mu_B$ at $\mu_S = \mu_Q = 0$, (b) $n_Q/T^3$ as a function of $T$ and $\mu_Q$ at $\mu_B = \mu_S = 0$, and (c) $n_S/T^3$ as a function of $T$ and $\mu_S$ at $\mu_B = \mu_Q = 0$.}
\label{fig:charges}
\end{figure}

The dimensionless conserved charges $n_B/T^3$, $n_Q/T^3$, $n_S/T^3$ are shown in Fig.~\ref{fig:charges} as a function of the temperature and respective chemical potential. The thermodynamic relation $n_B = (\partial P/\partial \mu_B)_{\mu_Q,\mu_S}$ is used for obtaining the net baryon density and similarly for the other variables. The net baryon density is almost linear in $\mu_B$ in the QGP phase in the limit where the other chemical potentials vanish as shown in Fig.~\ref{fig:charges} (a). The electric charge and strangeness densities exhibit a similar trend (Figs.~\ref{fig:charges} (b) and (c)). It should be noted that the conserved charges without the $1/T^3$ factor increase monotonically as a function of the temperature despite the peak structures near the pseudo-phase transition. They are also monotonic as a function of the respective chemical potential, \textit{e.g.} $\partial n_B/\partial \mu_B > 0$, which is required for thermodynamic stability.

\begin{figure}[tb]
\includegraphics[width=3.4in]{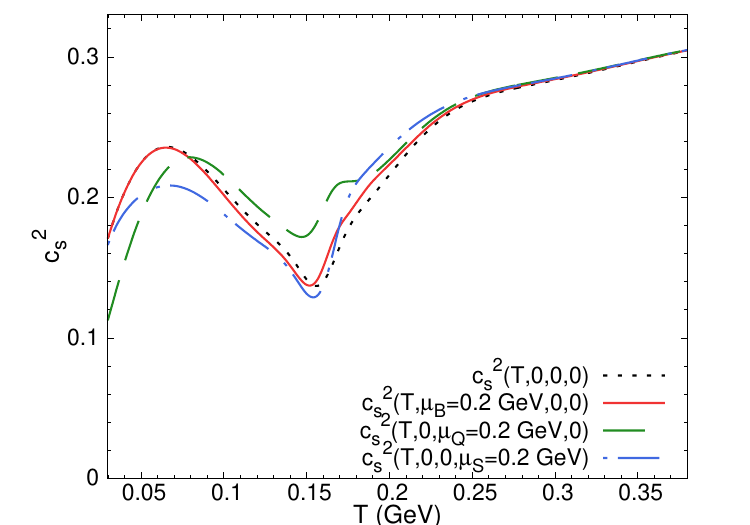}
\caption{(Color online) $c_s^2$ as a function of $T$ in the vanishing chemical potential limit and at $\mu_B = 0.2$ GeV, $\mu_Q = 0.2$ GeV, and $\mu_S = 0.2$ GeV, respectively.}
\label{fig:cs2}
\end{figure}

The squared sound velocity $c_s^2$ at fixed values of chemical potentials is shown in Fig.~\ref{fig:cs2}. Here $c_s^2$ is expressed as
\begin{align}
c_s^2 &= \left. \frac{\partial P}{\partial e} \right|_{n_B,n_Q,n_S} + \left. \frac{n_B}{e+P} \frac{\partial P}{\partial n_B} \right|_{e,n_Q,n_S}\nonumber \\
&= \left. \frac{n_Q}{e+P} \frac{\partial P}{\partial n_Q} \right|_{e,n_B,n_S} + \left. \frac{n_S}{e+P} \frac{\partial P}{\partial n_S} \right|_{e,n_B,n_Q} .
\end{align}
 It can be seen that the chemical potentials lead to non-trivial corrections of the sound velocity in the vicinity of the crossover and at lower temperatures. $\mu_Q$ has the largest effect and $\mu_B$ the least when the absolute value of the chemical potentials is the same, which is in line with the observation of the dimensionless pressure in the hadronic phase.

\begin{figure}[tb]
\includegraphics[width=3.4in]{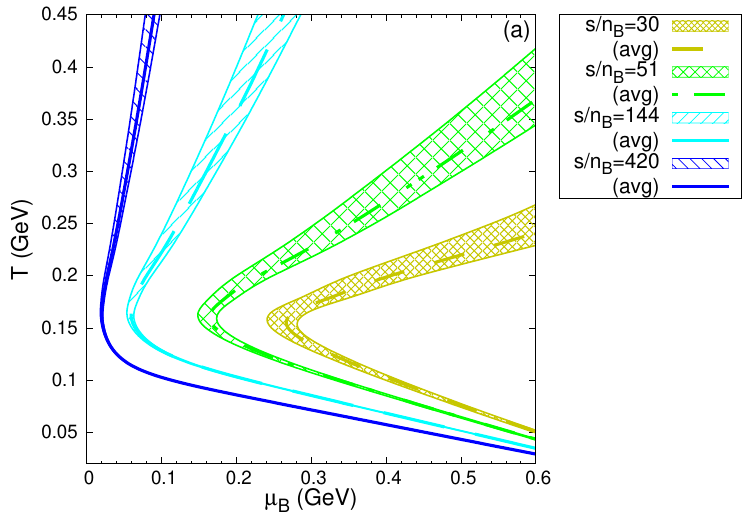}
\includegraphics[width=3.4in]{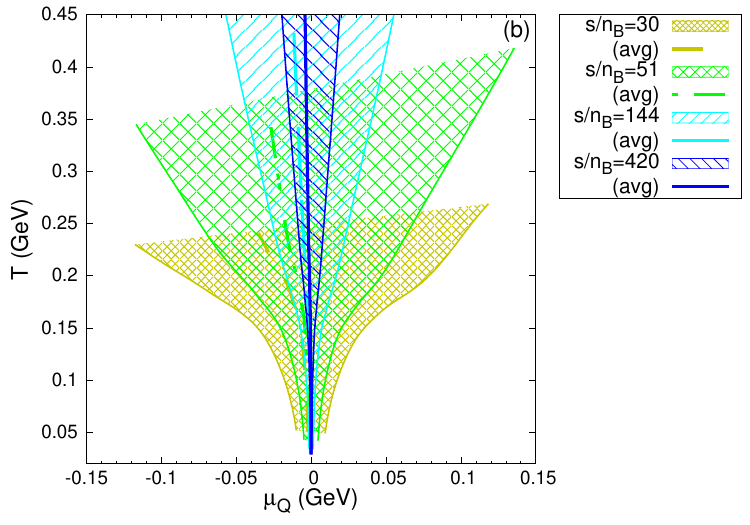}
\includegraphics[width=3.4in]{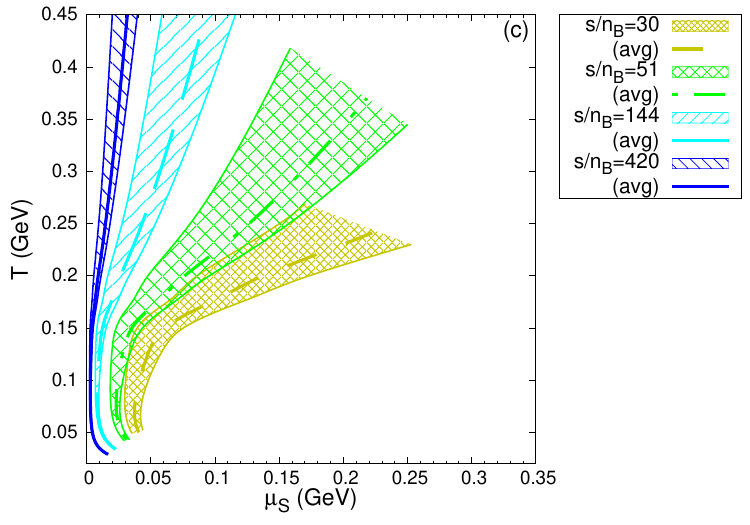}
\caption{(Color online) The regions explored when $n_Q/n_B$ ranges from 0 to 1 for $s/n_B$ = 420, 144, 51, and 30 on the (a) $T$-$\mu_B$, (b) $T$-$\mu_Q$, and (c) $T$-$\mu_S$ planes in the four-dimensional phase diagram. The dashed, dash-dotted, and dotted lines are the trajectories at $n_Q/n_B$ = 0.4. }
\label{fig:trajectories}
\end{figure}

Figure~\ref{fig:trajectories} illustrates the typical region in the phase space explored by nuclear collisions at different collision energies. Here, the constant $s/n_B$ lines illustrate approximate trajectories in nuclear collisions at given energies because the entropy and net baryon number are conserved during ideal hydrodynamic evolution. 
The trajectories are shown with bands, whose limits are defined by $n_Q/n_B = 0$ and 1. Here, the former corresponds to the regions with neutron-neutron sub-collisions and the latter to those with proton-proton sub-collisions caused by initial fluctuations in the positions of nucleons in colliding nuclei. It should be noted that the ratio can also be negative, for example when fluctuations and diffusion are considered. The strangeness neutrality condition $n_S/n_B = 0$ is imposed here. 

The average value of $n_Q/n_B$ in heavy nuclei such as gold or lead can be approximated to be 0.4 and is denoted by lines in the middle of the bands. We consider $s/n_B = 420, 144, 51$ and $30$, which correspond to the collider energies of $\sqrt{s_{\rm NN}}=200, 62.4, 19.6$ and $14.5$ GeV, to imitate the situations in the beam energy programs at RHIC \cite{Gunther:2016vcp}. The trajectory bands are truncated when the baryon chemical potential exceeds $\mu_B = 0.6$ GeV. One can see that the bands extend up to $|\mu_Q| \sim 0.1$ GeV and $\mu_S \sim 0.2$ GeV in the limit of $\mu_B = 0.6$ GeV at the lowest considered collision energies.

The trajectories bend near the crossover regions in the $T$-$\mu_B$ plane (Fig.~\ref{fig:trajectories} (a)) because $s/n_B \sim T/\mu_B$ holds at high temperatures in the QGP phase while a large $\mu_B$ is required to produce baryons with the mass over 1 GeV in the hadronic phase at lower temperatures. The trajectory of $n_Q/n_B = 0$ is about 25\% larger in the baryon chemical potential than that of $n_Q/n_B = 1$ in the QGP phase. This is consistent with the fact that $\mu_B = (5 n_B - n_Q)/T^2$ in the parton gas approximation (see also Sec.~\ref{sec3A} for a related discussion). On the other hand, distinguishing protons and neutrons would have a large effect in the exploration of the $T$-$\mu_Q$ plane, as shown in Fig.~\ref{fig:trajectories} (b). The average trajectory of $n_Q/n_B = 0.4$ leads to small negative values of $\mu_Q$, but the trajectory of neutron-rich domains $n_Q/n_B = 0$ probes larger negative values of $\mu_Q$ and that of proton-rich domains $n_Q/n_B = 1$ probes positive values of $\mu_Q$, resulting in wider bands. The bands are narrower in the hadronic phase because the electric charge is carried by pions, which can be relatively easily produced at small chemical potentials. 

Figure~\ref{fig:trajectories} (c) shows the trajectories in the $T$-$\mu_S$ plane. They behave similarly to the $T$-$\mu_B$ case with slightly wider trajectory bands; the chemical potential of an $n_Q/n_B = 0$ trajectory is twice that of the corresponding $n_Q/n_B = 1$ trajectory in the QGP phase. It can be understood in the parton gas approximation where $\mu_S = (2 n_B - n_Q)/T^2$ at higher temperatures. There is a second bending of trajectories at lower temperatures below $T \sim 0.05$ GeV, possibly because the kaon mass becomes non-negligible there. The results indicate that the four-dimensional equation of state can be essential for utilizing experimental data on nuclear collisions to explore the phase structure of QCD.
 
\section{Discussion}
\label{sec3}

\subsection{Conservation at particlization}

\begin{figure*}[tb]
\includegraphics[width=3.4in]{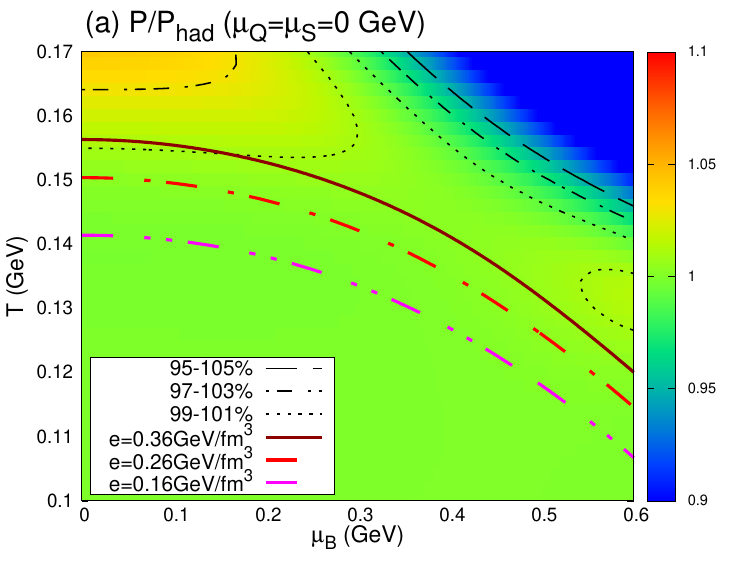}
\includegraphics[width=3.4in]{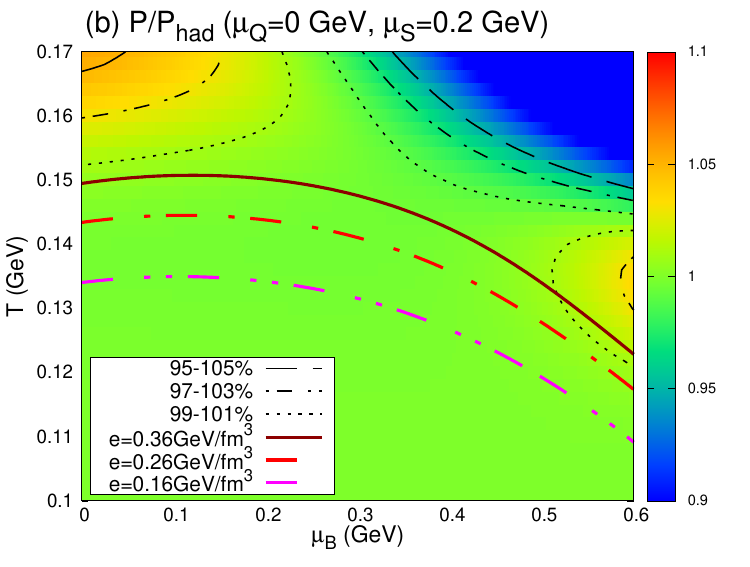}
\includegraphics[width=3.4in]{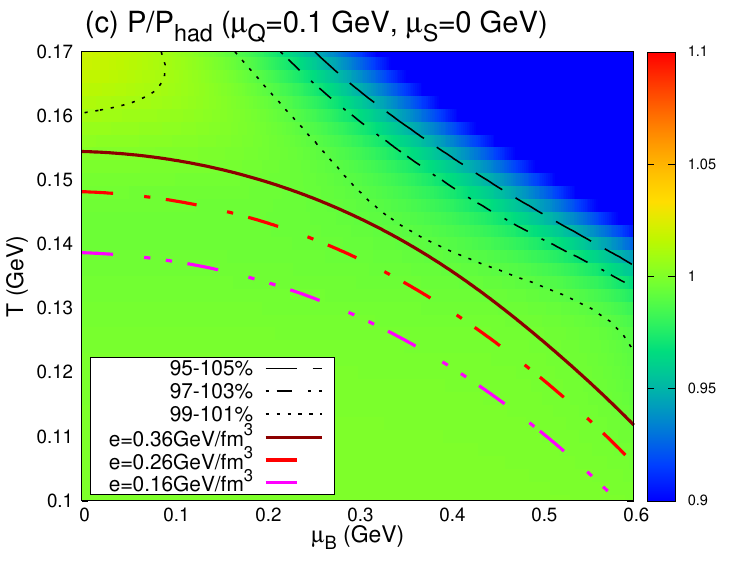}
\includegraphics[width=3.4in]{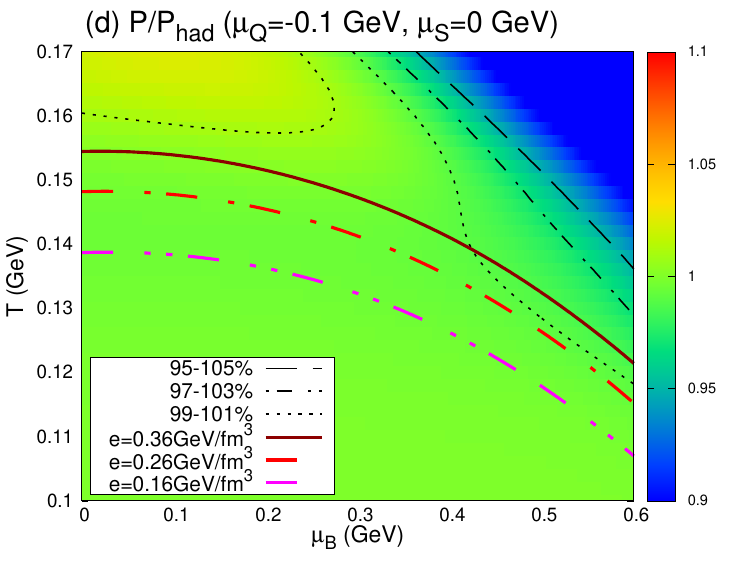}
\caption{(Color online) The ratio $P/P_\textrm{had}$ as a function of $T$ and $\mu_B$ at (a) $\mu_Q=\mu_S=0$ GeV, (b) $\mu_Q=0$ GeV, $\mu_S=0.2$ GeV, (c) $\mu_Q=0.1$ GeV, $\mu_S=0$ GeV, and (d) $\mu_Q=-0.1$ GeV, $\mu_S=0$ GeV. The thin dashed, dash-dotted, and dotted lines are the contours that denote the ranges where $|1-P/P_\textrm{had}|$ exceeds 5\%, 3\%, and 1\%, respectively. The thick solid, dash-dotted, and dash-double-dotted lines denote the constant energy density of 0.36, 0.26, and 0.16~GeV/fm$^{3}$, respectively.}
\label{fig:p_ratio}
\end{figure*}

The equation of state provides the relation among thermodynamic variables to close the set of equations in the hydrodynamic model of nuclear collisions. When the flow field is converted into particles, one often employs the Cooper-Frye formula \cite{Cooper:1974mv}, which is based on kinetic theory. It is required that the equation of state of the fluid match that of kinetic theory to ensure the conservation of energy, net baryon, strangeness, and electric charge at particlization for further simulations in a hadronic transport model. This implies that the particlization should occur in the domain of the phase diagram where the equation of state is dominantly described by the hadron resonance gas model in our approach.

The contour plots of the ratio $P/P_\mathrm{had}$ are shown in Fig.~\ref{fig:p_ratio} on the $T$-$\mu_B$ plane for four different sets of values of $\mu_Q$ and $\mu_S$: (a) $\mu_Q=\mu_S=0$ GeV, (b) $\mu_Q=0$~GeV and $\mu_S=0.2$~GeV, (c) $\mu_Q=0.1$~GeV and $\mu_S=0$~GeV, and (d) $\mu_Q=-0.1$~GeV and $\mu_S=0$~GeV to investigate regions that could be explored in nuclear collisions according to Fig.~\ref{fig:trajectories}. The lines of constant energy density are also plotted for $e = 0.36$, 0.26, and 0.16~GeV/fm$^{3}$ to illustrate typical switching energy densities from the hydrodynamic to transport models. It has been checked that the temperatures of those constant energy density lines are below the connecting temperature $T_c$ in Eq.~\eqref{eq:tc}.

If the particlization occurs where $P/P_\mathrm{had}$ is close to unity, the thermodynamic quantities will be conserved. One can see that the ratio becomes smaller than 1 in the QGP phase because the pressure would become exponentially large if there were no phase transition. On the other hand, the ratio is fairly close to 1 in the hadronic phase by construction. The difference between $P$ and $P_\mathrm{had}$ are within the 1\% range for all cases when $e = 0.26$ and 0.16~GeV/fm$^{3}$. When $e = 0.36$~GeV/fm$^{3}$, the difference is within the 3\% range. It should be noted that this is within the error bands of the original Lattice QCD simulations. 

\begin{figure*}[tb]
\includegraphics[width=2.3in]{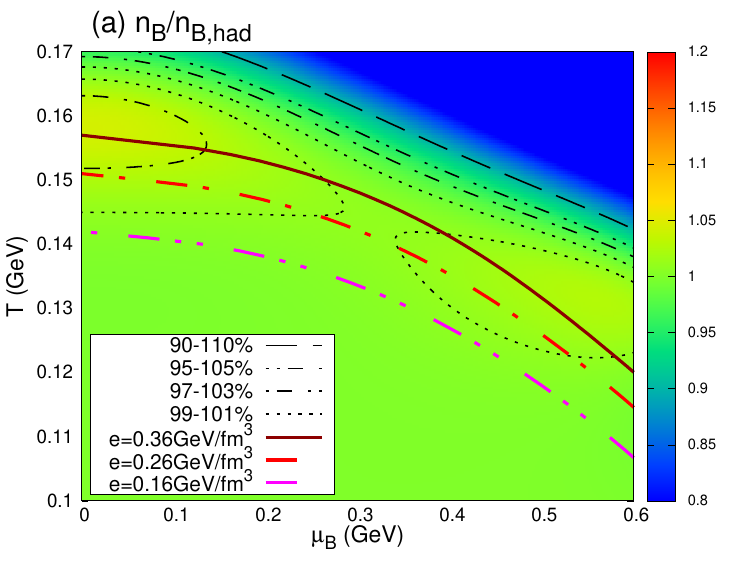}
\includegraphics[width=2.3in]{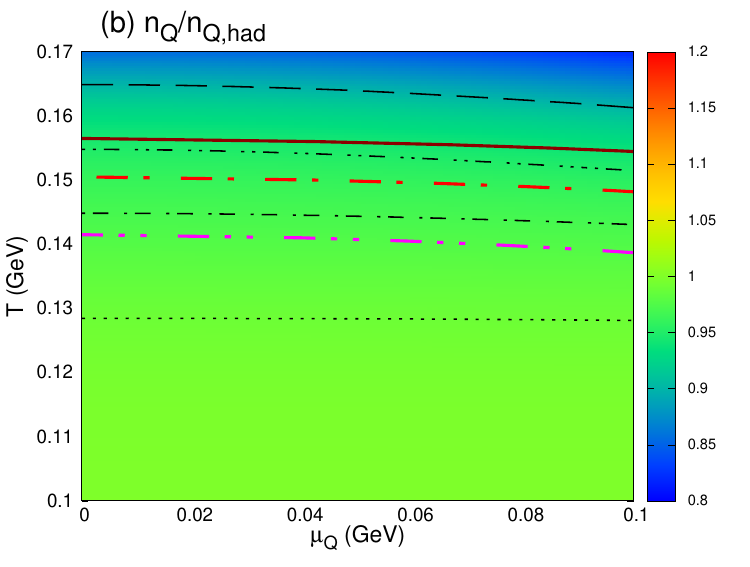}
\includegraphics[width=2.3in]{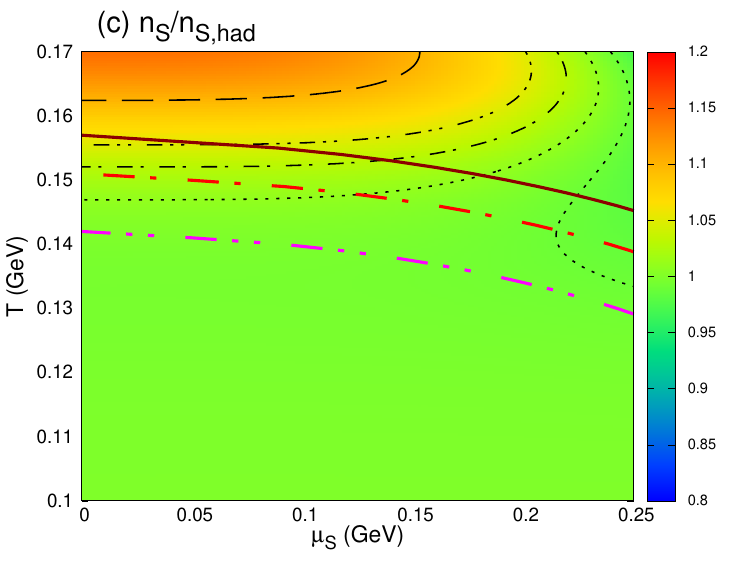}
\includegraphics[width=2.3in]{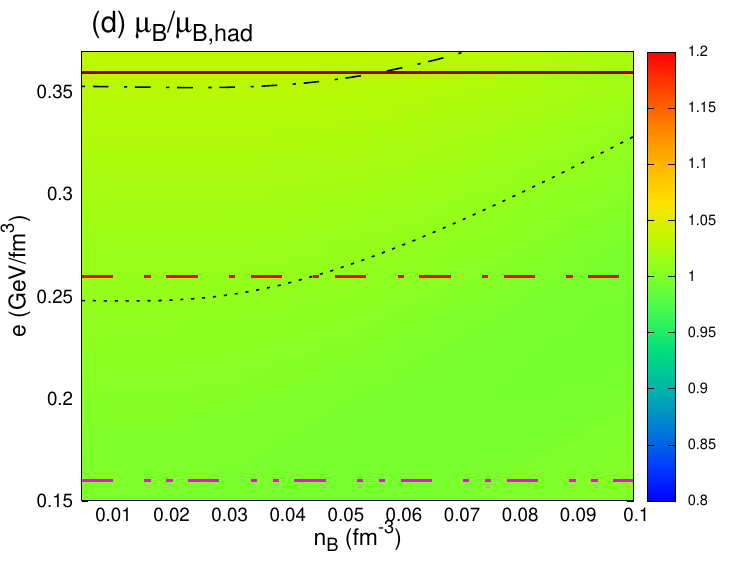}
\includegraphics[width=2.3in]{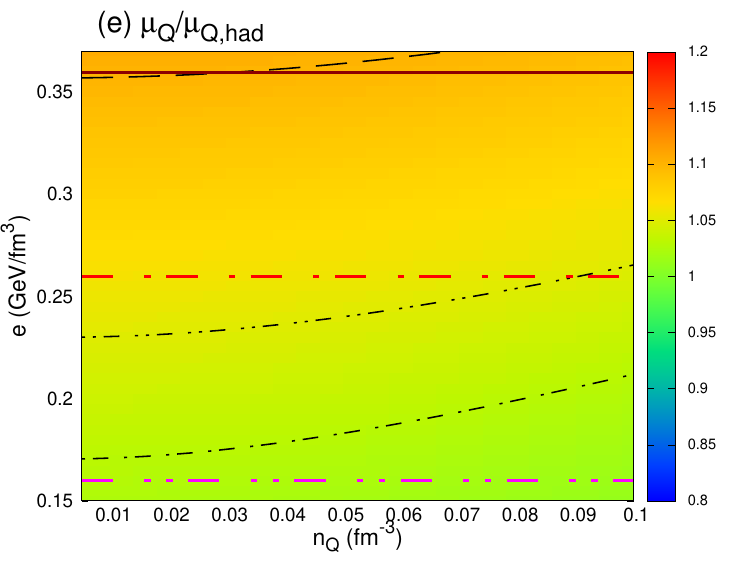}
\includegraphics[width=2.3in]{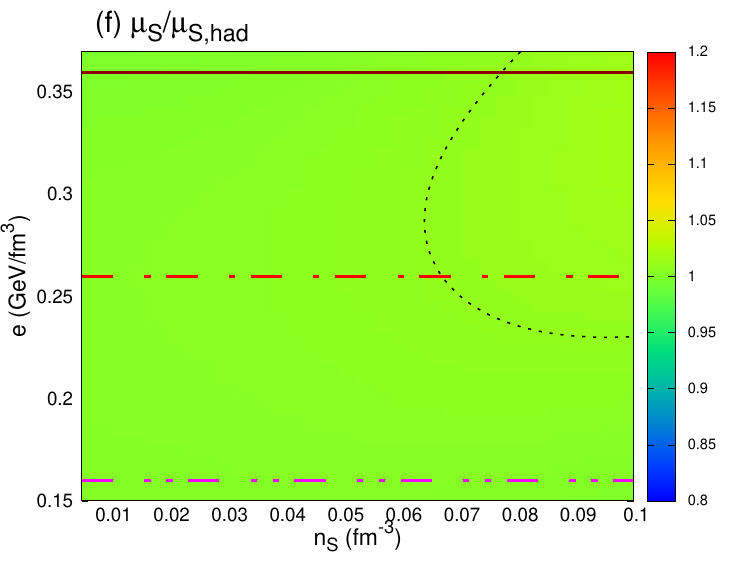}
\caption{(Color online) (a) $n_B/n_{B,\textrm{had}}$ as a function of $T$ and $\mu_B$ and (d) $\mu_B/\mu_{B,\textrm{had}}$ as a function of $e$ and $n_B$ at $\mu_Q=\mu_S=0$, (b) $n_Q/n_{Q,\textrm{had}}$ as a function of $T$ and $\mu_Q$ and (e) $\mu_Q/\mu_{Q,\textrm{had}}$ as a function of $e$ and $n_Q$ at $\mu_B=\mu_S=0$, and (c) $n_S/n_{S,\textrm{had}}$ as a function of $T$ and $\mu_S$ and (f) $\mu_S/\mu_{S,\textrm{had}}$ as a function of $e$ and $n_S$ at $\mu_B=\mu_Q=0$.}
\label{fig:n_ratio}
\end{figure*}

On the other hand, net baryon, electric charge, and strangeness densities are derivatives of the pressure and can pose more stringent constraints on the switching energy densities. We consider the ratios of the conserved charges $n_B, n_Q$, and $n_S$ to those in the resonance gas model $n_{B,\mathrm{had}}$, $n_{Q,\mathrm{had}}$, and $n_{S,\mathrm{had}}$. The top panels of Fig.~\ref{fig:n_ratio} show $n_B/n_{B,\textrm{had}}$ as a function of $T$ and $\mu_B$, $n_Q/n_{Q,\textrm{had}}$ as a function of $T$ and $\mu_Q$, and $n_S/n_{S,\textrm{had}}$ as a function of $T$ and $\mu_S$  in the vanishing limit of the other chemical potentials. When $e=0.36$ GeV/fm$^3$, the deviation can reach the 3-5\% range in the case of net baryon, the 5-10\% range in the cases of net charge and strangeness. The deviation is within 3\% when $e=0.16$ GeV/fm$^3$. We also consider the ratios of the conjugate variables $\mu_B$, $\mu_Q$, and $\mu_S$ to their counterparts in the hadron resonance gas model to illustrate the situation at particlization. The results are shown in the bottom panels of Fig.~\ref{fig:n_ratio}. They cover the ranges of chemical potentials that could be relevant to nuclear collisions; $\mu_B = 0.602$~GeV at $e=0.15$~GeV/fm$^3$ and $n_B = 0.1$~fm$^{-3}$ in Fig.~\ref{fig:n_ratio} (d), $\mu_Q = 0.114$~GeV at $e=0.15$~GeV/fm$^3$ and $n_Q = 0.1$~fm$^{-3}$ in Fig.~\ref{fig:n_ratio} (e), and $\mu_S = 0.238$~GeV at $e=0.15$~GeV/fm$^3$ and $n_S = 0.1$~fm$^{-3}$ in Fig.~\ref{fig:n_ratio} (f). One can see that the deviation of $\mu_B/\mu_{B,\textrm{had}}$ as a function of $e$ and $n_B$ at $\mu_Q=\mu_S=0$ is in the 3-5\% range for $e=0.36$ GeV/fm$^3$. Likewise, that of $\mu_Q/\mu_{Q,\textrm{had}}$ as a function of $e$ and $n_Q$ at $\mu_B=\mu_S=0$ is in the 5-10\% range and that of $\mu_S/\mu_{S,\textrm{had}}$ as a function of $e$ and $n_S$ at $\mu_B=\mu_Q=0$ in the 1-3\% range. The results indicate that particlization at the lowest energy is preferred for consistency with the kinetic theory. It should be noted that the maximum deviation could be larger when an entire four-dimensional space is explored. The agreement for higher switching energy densities might be improved by future implementation of the $\mu_Q$ and $\mu_S$ dependencies of the connecting temperature. 

\subsection{Dependence on hadronic components}

\begin{figure*}[tb]
\includegraphics[width=3.4in]{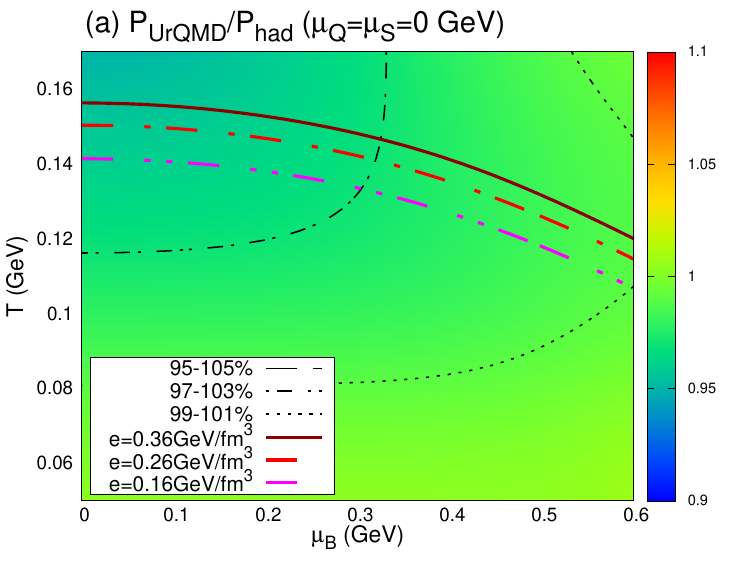}
\includegraphics[width=3.4in]{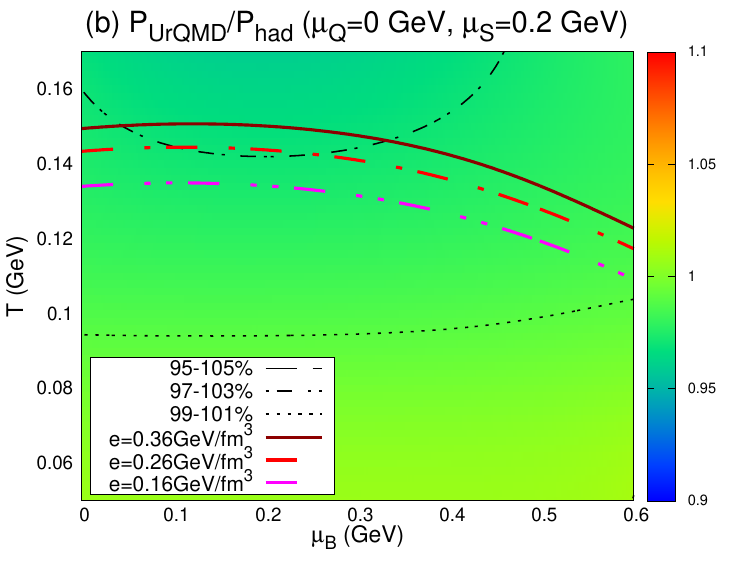}
\includegraphics[width=3.4in]{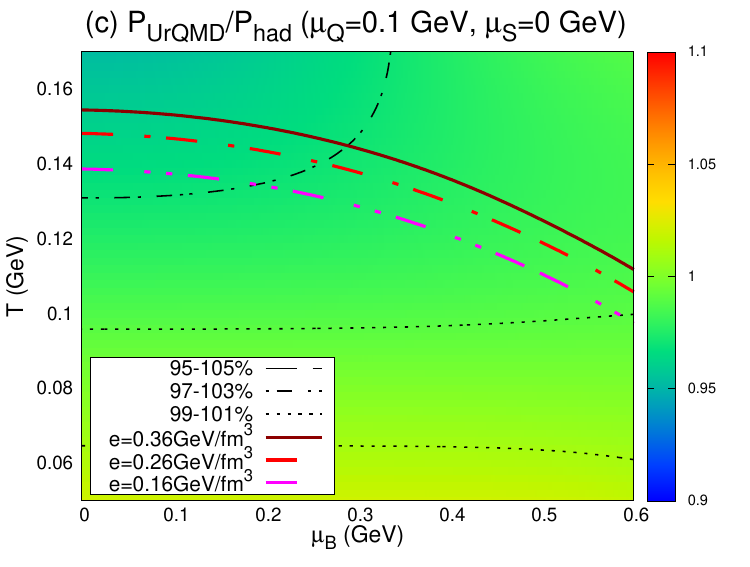}
\includegraphics[width=3.4in]{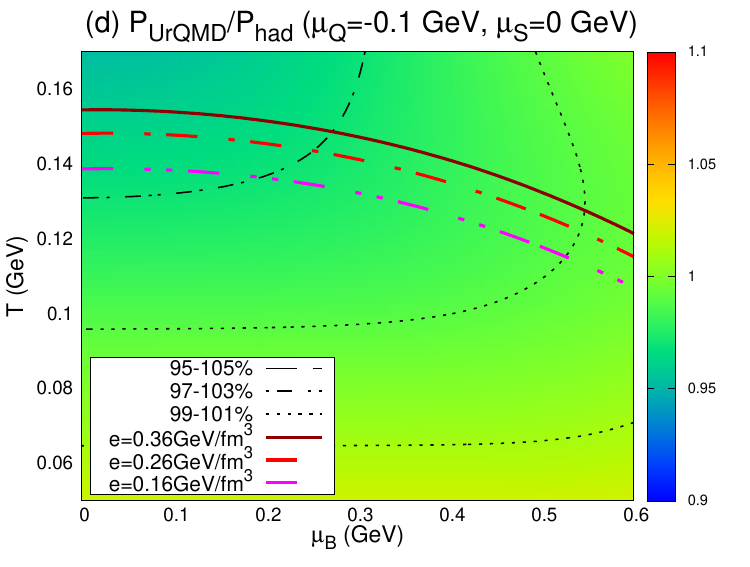}
\caption{(Color online) The ratio $P_\textrm{UrQMD}/P_\textrm{had}$ as a function of $T$ and $\mu_B$ at (a) $\mu_Q=\mu_S=0$ GeV, (b) $\mu_Q=0$ GeV, $\mu_S=0.2$ GeV, (c) $\mu_Q=0.1$ GeV, $\mu_S=0$ GeV, and (d) $\mu_Q=-0.1$ GeV, $\mu_S=0$ GeV. The thin dashed, dash-dotted, and dotted lines are the contours that denote the ranges where $|1-P_\textrm{UrQMD}/P_\textrm{had}|$ exceeds 5\%, 3\%, and 1\%, respectively. The thick solid, dash-dotted, and dash-double-dotted lines denote the constant energy density of 0.36, 0.26, and 0.16~GeV/fm$^{3}$, respectively.}
\label{fig:urqmd}
\end{figure*}

We next investigate the validity of using a transport model that has a different content of hadronic resonances compared to the model used in constructing the equation of state. We compare the hadronic pressure based on the list of particles from our model, $P_\mathrm{had}$, with that based on the list from the latest version of the Ultra-relativistic-Quantum-Molecular-Dynamics (UrQMD) model \cite{Bass:1998ca,Bleicher:1999xi}, $P_\mathrm{UrQMD}$. Comparison to other models such as SMASH \cite{SMASH:2016zqf} is left for future studies.

The ratio $P_\mathrm{UrQMD}/P_\mathrm{had}$ is shown for the four sets of different values of $\mu_Q$ and $\mu_S$ in Fig.~\ref{fig:urqmd}. One can find about 1-5\% difference in the pressure along the lines of the switching energy densities at $e = 0.36, 0.26$, and 0.16~GeV/fm$^{3}$ in all cases.
This difference aligns with the results at $\mu_B = \mu_Q = \mu_S = 0$ investigated in Ref.~\cite{JETSCAPE:2020mzn}.
The dependencies of the ratio on $\mu_Q$ and $\mu_S$ are also found to be small. The results nevertheless imply that the matching of hadronic components can be important for precision analyses.

\subsection{Tabulation method}
\label{sec3A}

\begin{figure*}[tb]
\includegraphics[width=3.4in]{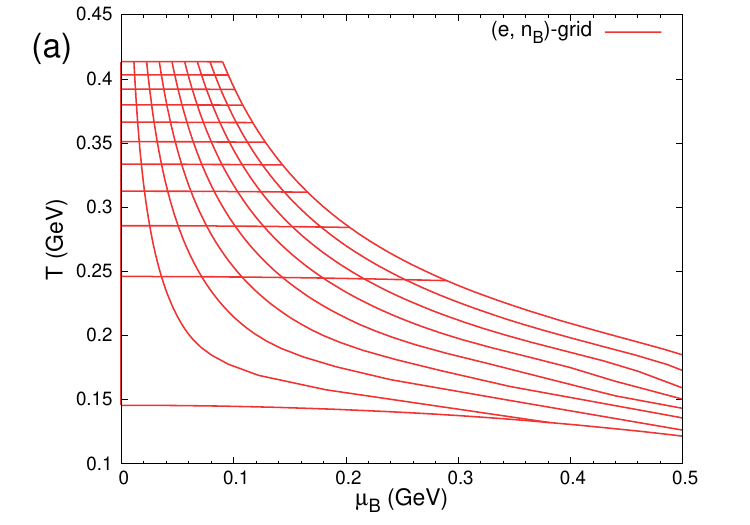}
\includegraphics[width=3.4in]{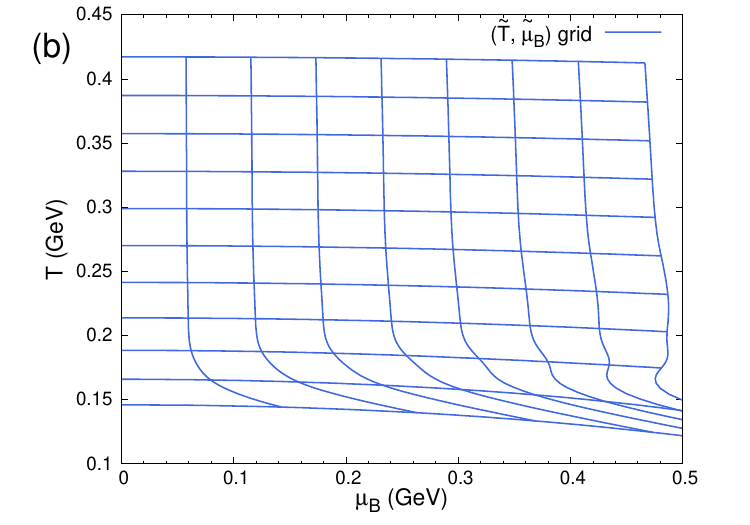}
\caption{(Color online) Comparison of grids with equal spacing in (a) $e$ and $n_B$ and (b) $\tilde{T}$ and $\tilde{\mu_B}$ in the $T$-$\mu_B$ plane.}
\label{fig:grids}
\end{figure*}

The equation of state in numerical hydrodynamic models for nuclear collisions is often tabulated for numerical efficiency. The data points are interpolated to obtain values in between. This has been established to be a valid method when one or two variables, such as the energy density and the net baryon number, are considered. For higher-dimensional equations of state, the number and size of tables can become too large for regular numerical implementation. 
This mainly comes from the fact that the energy density increases roughly as $T^4$ and the conserved charge densities as $\mu T^2$ where $\mu$ is an arbitrary chemical potential. 

A two-dimensional example case of the grid with equal spacing in $e$ and $n_B$ on the $T$-$\mu_B$ plane is illustrated at $\mu_Q=\mu_S=0$ GeV in Fig.~\ref{fig:grids} (a). The ranges are set to $0.2 \leq e \leq 50.2$ GeV/fm$^{3}$ and $0 \leq n_B \leq 0.64$ fm$^{-3}$ and the spacing to $\Delta e = 5$ GeV/fm$^3$ and $\Delta n_B = 0.08$ GeV/fm$^3$ to partially cover the crossover region. One can see that the grid is warped owing to the non-linearity of the thermodynamic relations. Increasing the number of data points in lower temperature regions to validate the interpolation would lead to the redundancy of data points in higher temperature regions. The inefficiency can be dealt with by introducing multiple tables to change the grid spacing and allowing the inevitable redundancy in the two dimensional case, but this method becomes increasingly difficult in the higher dimensional cases.

To overcome the problem, we introduce variables $\tilde{T}$, $\tilde{\mu}_B$, $\tilde{\mu}_Q$, and $\tilde{\mu}_S$, which are the temperature and chemical potentials of the parton gas with the same energy and conserved charge densities, for constructing the tables of the equation of state. They are implicitly defined in the parton gas equation of state with $N_f = 3$ as
\begin{align}
    e &= \frac{19\pi^2}{12} \tilde{T}^4 , \label{eq:e2}\\
    n_B &= \frac{1}{3} \tilde{\mu}_B \tilde{T}^2 - \frac{1}{3} \tilde{\mu}_S \tilde{T}^2, \\
    n_Q &= \frac{2}{3} \tilde{\mu}_Q \tilde{T}^2 + \frac{1}{3} \tilde{\mu}_S \tilde{T}^2, \\
    n_S &= - \frac{1}{3} \tilde{\mu}_B \tilde{T}^2 + \frac{1}{3} \tilde{\mu}_Q \tilde{T}^2 + \tilde{\mu}_S \tilde{T}^2 ,\label{eq:ns2}
\end{align}
whose solutions can be obtained analytically as
\begin{align}
    \tilde{T}(e,n_B,n_Q,n_S) &= \bigg(\frac{12}{19\pi^2} e \bigg)^{1/4} ,\label{eq:pg-t} \\ 
    \tilde{\mu}_B(e,n_B,n_Q,n_S) &= \frac{5 n_B-n_Q+2n_S}{\tilde{T}^2}, \label{eq:pg-b}\\
    \tilde{\mu}_Q(e,n_B,n_Q,n_S) &= \frac{- n_B+2n_Q-n_S}{\tilde{T}^2}, \label{eq:pg-q}\\
    \tilde{\mu}_S(e,n_B,n_Q,n_S) &= \frac{2 n_B - n_Q + 2n_S}{\tilde{T}^2}.\label{eq:pg-s}
\end{align}

The equation of state can be constructed and tabulated as a function of those variables as $P = P(\tilde{T},\tilde{\mu}_B,\tilde{\mu}_Q,\tilde{\mu}_S)$ using Eqs.\,\eqref{eq:e2}-\eqref{eq:ns2} and thermodynamic relations in the whole region of temperatures and chemical potentials, which reduces the size of the table. A grid with equal spacing in $\tilde{T}$ and $\tilde{\mu}_B$ is shown as a two-dimensional example case at $\mu_Q=\mu_S=0$ GeV in the $T$-$\mu_B$ plane for comparison (Fig.~\ref{fig:grids} (b)). It covers the ranges of $0.1 \leq \tilde{T} \leq 0.4$ GeV and $0 \leq \tilde{\mu}_B \leq 0.48$ GeV with $\Delta \tilde{T} = 0.03$ GeV and $\Delta \tilde{\mu}_B = 0.06$ GeV. It can be found that the grid has almost equal spacing in $T$ and $\mu_B$ in the QGP phase towards higher temperatures as the deviation from the parton gas equation of state becomes smaller. The grid, on the other hand, is warped in the hadronic phase, though to a lesser extent than that of $e$ and $n_B$ because the dimensions of the variables are the same.

The thermodynamic variables that appear in hydrodynamic equations of motion, $e$, $n_B$, $n_Q$, and $n_S$, can be analytically converted into the pseudo-temperatures and chemical potentials using Eqs.\,\eqref{eq:pg-t}-\eqref{eq:pg-s} to access the table. See Ref.~\cite{Pihan:2023dsb,Pihan:2024lxw} for practical utilization of the method for using \textsc{neos-4d} in a hydrodynamic model. A different method of implementation can be found in Ref.~\cite{Plumberg:2023vkw,Plumberg:2024leb}.

\section{Conclusions}
\label{sec4}

We have developed a four-dimensional model of the QCD equation of state with multiple chemical potentials corresponding to the conserved charges in relativistic nuclear collisions: net baryon, electric charge, and strangeness. It can be used for hydrodynamic modeling of relativistic nuclear collisions over a wide range of energies and of different nuclear species. The latest lattice QCD results of the pressure and susceptibilities calculated at zero densities are used to construct a finite-density equation of state in the QGP phase. A crossover-type equation of state is constructed by smoothly matching the pressure to that of the hadron resonance gas model in the hadronic phase. Thermodynamic consistency has been numerically verified within the parameter space of consideration.

Without the traditional constraint of $n_Q/n_B \sim 0.4$, a wider region of the phase diagram can be explored in the beam energy scan programs, including large positive and negative regions of $\mu_Q$ in the analyses of relativistic heavy-ion collisions, because the ratio $n_Q/n_B$ can vary from 0 to 1 in neutron-rich and proton-rich domains of the hot QCD matter. 

The conservation at particlization has been investigated to check the validity of the numerical implementation of the obtained equation of state to hydrodynamic models. The difference between the pressure of the \textsc{neos-4d} model and that of the hadron resonance gas is found to be within the 3\% range when the switching energy density is $e=0.36$ GeV/fm$^{3}$ and within the 1\% range when $e=0.16$ and 0.26 GeV/fm$^{3}$. The agreement with kinetic theory prefers particlization at lower energy densities. It should, on the other hand, be noted that reproduction of experimental data in the hydrodynamic model can require onset of hadronic transport at higher energies. Since this requirement is dependent on the collision system, it is difficult to determine a single set of optimal values for the particlization temperature/chemical potentials.

We also compared the pressure to the pressure from the slightly different UrQMD particle list, finding that the difference is about 1-5\% for the aforementioned energy densities. This implies that the particle lists are mostly consistent and different hadronic afterburners might be used. It is nevertheless recommended to match exactly the particle lists of the equation of state and the transport model for precision analyses. 

We have also developed an efficient numerical method for using a multi-dimensional QCD equation of state in hydrodynamic simulations by introducing the pseudo-temperatures and chemical potentials $\tilde{T},\tilde{\mu}_B,\tilde{\mu}_Q$, and $\tilde{\mu}_S$ for the construction of the pre-calculated tables that can be used in hydrodynamic models. These are defined as the temperature and the chemical potentials of the parton gas for the given energy and conserved charge densities.

The equation of state developed in this work is a necessary ingredient to study in all generality nuclear collisions over a wide range of beam energies and for different nuclei based on hydrodynamic approaches. It further allows to explore effects of initial fluctuations \cite{Martinez:2019jbu,Carzon:2019qja,Carzon:2023zfp} and diffusion of conserved charges \cite{Greif:2017byw,Fotakis:2019nbq}, which can further broaden the region of exploration in the phase diagram. Interesting future venues for progress are the improvement of the equation of state by introducing the $\mu_Q$ and $\mu_S$ dependence of the connecting temperature as well as by inclusion of the critical ``plane" and the first order phase transition. Also, novel methods of lattice QCD-based estimation of the finite-density equation of state \cite{Borsanyi:2021sxv,Borsanyi:2022qlh,Mondal:2021jxk,Mukherjee:2021tyg,Mitra:2022vtf} could be introduced in the modeling.

The four-dimensional version of the QCD equation of state model \textsc{neos} is 
publicly available \cite{neos4d}.

\begin{acknowledgments}
The authors thank Frithjof Karsch, Swagato Mukherjee, and Sayantan Sharma for providing the lattice QCD data. This work is supported by JSPS KAKENHI Grant Numbers JP19K14722 and JP24K07030 (A.M.), by the U.S. Department of Energy, Office of Science, Office of Nuclear Physics, under DOE Contract No.~DE-SC0012704 and within the framework of the Saturated Glue (SURGE) Topical Theory Collaboration (B.P.S.) and Award No.~DE-SC0021969 (C.S. \& G.P.). C.S. acknowledges a DOE Office of Science Early Career Award. 
\end{acknowledgments}

\bibliography{nxeos}

\end{document}